# Real World Scrum
# A Grounded Theory of Variations in Practice

Zainab Masood, Rashina Hoda, and Kelly Blincoe

**Abstract**—Scrum, the most popular agile method and project management framework, is widely reported to be used, adapted, misused, and abused in practice. However, not much is known about how Scrum actually works in practice, and critically, where, when, how and why it diverges from Scrum by the book. Through a Grounded Theory study involving semi-structured interviews of 45 participants from 30 companies and observations of five teams, we present our findings on how Scrum works in practice as compared to how it is presented in its formative books. We identify significant variations in these practices such as work breakdown, estimation, prioritization, assignment, the associated roles and artefacts, and discuss the underlying rationales driving the variations. Critically, we claim that not all variations are process misuse/abuse and propose a nuanced classification approach to understanding variations as *standard, necessary, contextual,* and *clear deviations* for successful Scrum use and adaptation.

---◆---

## 1   INTRODUCTION

AGILE software development methods such as Scrum [1], [2] and XP [4]-[6] follow a collaborative, people-oriented approach to software development and embody the core Agile Manifesto values[7]. Scrum is by far the most popular and commonly used agile method [8]. It is practiced by many large and small companies to varying degrees (e.g., Yahoo!, Microsoft, and Google) [2]. Scrum is an iterative and incremental method focusing on project management practices [9]. Its key practices such as estimation, breakdown, and prioritization primarily focus on work planning [1], [2] while the practice of self-assignment is considered a hallmark feature of self-organizing teams [28]. In essence, it revolves around organising people and providing procedures to add business value and deliver quality through effective planning [10].

Perhaps, one of the reasons behind Scrum's dominance [8] is its perceived simplicity and "lightweight" approach to managing software projects as described in its formative literature such as the Scrum Guide and the Scrum Primer [1], [2] (referred to in this paper as 'Scrum by the book'). These concise guides provide an easy to understand overview of Scrum's practices, roles, and artefacts.

The perceived simplicity of Scrum by the book is corroborated by Scrum enthusiasts who claim teams must adhere to its practices in their entirety and "by the book" to avail the real benefits [15]. Ken Schwaber, a co-creator of Scrum suggests teams customize it to suit their *'dysfunctions'* or *'inadequacies'* and states *"I estimate that 75% of those organizations using Scrum will not succeed in getting the benefits that they hope for from it"* [17]. The other co-creator, Jeff Sutherland, is equally skeptical of variations and labels these deviations from the recommended Scrum practices as "ScrumButts" [33].

Yet, they also acknowledge that Scrum can be *"difficult to master"* [1]. This could be because Scrum by the book is not prescriptive about its key project management practices (including breakdown, estimation, prioritization, sprint goal creation, refinement, and work assignment), leaving implementation details to individual practitioners. It is no surprise then that many organisations are seen to use Scrum variants or modify it to suit their settings [12]-[14]. Research shows contextual adaptations can be necessary and beneficial [16],[18],[19].

While several studies have reported variations observed in practice, these have mostly been identified as secondary findings in studies with another primary focus, e.g. understanding the daily standup [23] or the product backlog [22]. A limited number of studies have exclusively focused on variations, reporting only *preliminary findings* [12], [13], or on a specific role, e.g. the Product Owner (PO) [11]. Thus far, no study presents descriptive and nuanced research on Scrum variations, grounded in substantial and detailed qualitative evidence from practice.

Our Grounded Theory study was guided by the following research question: **How, when and why does Scrum practice vary from Scrum by the book?** Based on semi-structured interviews of 45 practitioners and observations of five teams, we present variations in key Scrum roles, artefacts, and project management practices (including breakdown, estimation, prioritization, sprint goal creation, refinement, and work assignment). Critically, we describe how, when and why these variations occur and propose a nuanced classification approach to making sense of variations in practice.

- *Zainab Masood is with the Department of Electrical, Computer, and Software Engineering, University of Auckland, New Zealand.*
  *E-mail: zmas690@aucklanduni.ac.nz.*
- *Rashina Hoda is with the Faculty of Information Technology, Monash University, Melbourne, Australia. E-mail: rashina.hoda@monash.edu.*
- *Kelly Blincoe is with the Department of Electrical, Computer, and Software Engineering, University of Auckland, New Zealand.*
  *E-mail: k.blincoe@auckland.ac.nz*

## 2 RELATED WORK

Prior research has investigated various development methodologies, their adoption and adaptation due to different interpretations, organizational constraints [40]-[42] in different fields. In software engineering, researchers have explored how agile methods are tailored to meet organizational or project needs in practice [16], [18], [19]. *"The kinds of projects that the method designers had in mind when they constructed the first Agile methods"* are termed the agile sweet spot [34]. Differences between the sweet spot and other contexts and constraints are seen to necessitate adaptations [11],[16],[20]. Of the popular agile methods, XP has been found to be surprisingly resistant to adaptation or tailoring [19] reporting less number of studies identifying variations [44], perhaps explaining in part its gradual decline in industry adoption over the years [3] from a reported 23% in 2007 [35] to 1% in 2019 [8].

Adaptations to Scrum, on the other hand, have been widely addressed and largely criticized by Scrum evangelists [17],[33] and recorded in research literature [11],[12],[20]-[23]. Several studies mention or touch upon variations as part of a related or different study focus [20]-[23], [43]. These include statistical surveys of agile adoption that also reported on method compliance [21] and adaptation [20]. Kurapati et al. reported one third of their 109 survey respondents were fully compliant, nearly half were strongly compliant, and the remaining reported weak (12%) and no compliance (9%) with Scrum [21]. Details and examples of which specific variations occurred and the potential rationales behind them were not reported. Another adoption survey in 2011 [20] reported variations in Scrum roles such as the presence of project managers alongside Scrum Masters, [37],[8]. Again, the motivations and rationales behind the observed figures were not reported [20].

More recent qualitative studies describe some variations in practice as part of a different study focus, e.g. a specific Scrum practice [23] or artefact [22] or as part of the agile transformational journey in a distributed setting [43]. Based on a comprehensive Grounded Theory study of the Scrum practice, the daily standup, *regular*, instead of *daily*, standup was recommended as a common variation. Additionally, the primary purpose of the meeting defined in the Scrum Guide as "synchronization of activities and planning" [1] was not seen to be supported in practice. An investigation into the generation and role of the Scrum artefact, the *product backlog*, also revealed some variations in its use in practice such as partial ordering and items lacking estimation and details [22]. In a case study on agile transformation, Lous et al. found some adaptions of Scrum [43]. Rationales were identified for some of the variations (e.g. skipping sprint planning meetings as the content of these meetings was not of interest to the entire team).

Few studies focus exclusively on the topic of Scrum variations and most provide only preliminary findings [11], [12], [13]. One investigated variations in a single Scrum role, the Product Owner (PO) [11]. Based on interviews with five active POs, an observed variation was the PO's actual availability on the projects varied compared to the recommended easy accessibility in Scrum by the book.

A critical study of Scrum variations identified 14 anti-patterns or *"potentially harmful practices"* e.g. big requirements document, PO without authority, and no sprint retrospective [13]. Based on data from 18 Finnish software practitioners, it acknowledged that some anti-patterns are justified in specific cases. The study set an agenda for future in-depth studies in wider contexts.

Diebold et al. conducted an investigation into Scrum adaptations across a range of aspects including sprint lengths, team size, requirements engineering, and quality assurance [12]. Based on 10 interviews conducted in German companies, contrary to [21], compliance to the Scrum Guide was found to be low. Reported reasons behind variations included perceived efficiency and legacy habits from traditional ways of working.

Unlike statistical surveys [20], [21], our qualitative study aims to answer not just what variations occur, but how, when and why they occur in practice. Unlike studies reporting secondary findings on variations as part of different study focus [22], [23], this paper focuses exclusively on variations. Unlike [12], [13], our results extend beyond preliminary findings.

Critically, our study presents a *descriptive theory of Scrum variations* in its key project management practices, grounded in practice. Since Scrum primarily focuses on project management, it is important to understand how, when and why Scrum project management practices (including breakdown, estimation, assignment, prioritization, and sprint goal creation) vary from Scrum by the book.

## 3 RESEARCH METHODS

We applied the Grounded Theory (GT) method for data collection, analysis, and reporting our findings [24], [25], [32]. As the study aimed to understand and investigate prescribed methods and associated practice variations, GT was well-suited to our aims as it enables the investigation of real-world phenomenon as well as comparison across multiple sources of information, in this case, across findings from studying real-world Scrum teams (in practice) and seminal Scrum guides (by the book). Our study adds to the growing body of agile literature using GT [16], [22], [23], [26]-[29]. We employed the *Strauss-Corbinian* version of GT due to its prescriptive and structured approach to data analysis, and easy to follow guidelines.

### 3.1 Data Collection
We collected data from two main data sources, industrial data (in practice) and the basic Scrum guides (by the book):

### 3.1.1 By the Book
The Scrum Guide [1] supplemented by the Scrum Primer [2], the formative Scrum texts, were used as the data source to understand what is prescribed in Scrum *by book*. These are are commonly acknowledged and referenced in research studies as the fundamental Scrum references, the Guide being cited as the definitive source [11], [12], [13], [18], [19], [20], [22], [23].

TABLE 1 PARTICIPANT AND TEAM OBSERVATIONS

**INTERVIEWS**

| P# | ROLE | DOMAIN | TX | AX |
|---|---|---|---|---|
| P1 | Team Lead | Info Tech | 11 | 6-7 |
| P2 | Software Engineer | Info Tech | 1 | 2.5 |
| P3 | Associate Team Lead | Info Tech | 4-5 | 4-5 |
| P4 | Software Engineer | Info Tech | 2.5 | 2.5 |
| P5 | Team Lead | Info Tech | 7 | 7 |
| P6 | Senior Software Engineer | Info Tech | 4 | 2 |
| P7 | Team Lead | Info Tech | 7.5 | 7.5 |
| P8 | Product Owner | Telecom | 12 | 5 |
| P9 | Consultant | Info Tech | 10 | 3 |
| P10 | Team Lead | Medical | 13 | 7 |
| P11 | Developer; Scrum Master | Transport | 17 | 7 |
| P12 | Developer | Info Tech | 10 | 6 |
| P13 | Developer | Accounting | 2 | 2 |
| P14 | Senior Architect | IC Tech | 10 | 3 |
| P15 | Test Analyst | Finance | 10 | 5 |
| P16 | Tester | Medical | 12 | 1 |
| P17 | Developer; Scrum Master | Info Tech | 8 | 3.5 |
| P18 | Lead Developer | Info Tech | 25 | 9 |
| P19 | Developer; Scrum Master | Info Tech | 12 | 7 |
| P20 | Developer | Info Tech | 4 | 3.5 |
| P21 | Development Manager | Info Tech | 14 | 9 |
| P22 | Developer | Medical | 2.5 | 1.5 |
| P23 | Development Manager | Medical | 20 | 2 |
| P24 | Lead Developer | Medical | 20 | 3 |
| P25 | Scrum Master | Medical | 9 | 6 |
| P26 | Developer | Medical | 12.5 | 6 |
| P27 | Tester | Medical | 10 | 3 |
| P28 | Developer | Medical | 12 | 2 |
| P29 | Developer | Medical | 10.5 | 4 |
| P30 | Head of Product Delivery | Healthcare | 13 | 3 |
| P31 | Developer Consultant | Retail | 10 | 5 |
| P32 | Tester | Info Tech | 5 | 3 |
| P33 | Consultant | Info Tech | 11 | 4 |
| P34 | Senior Architect | Info Tech | 15 | 10 |
| P35 | Tester | Finance | 16 | 14 |
| P36 | Quality Assurance Analyst | Finance | 7.5 | 2.5 |
| P37 | Scrum Master | Info Tech | 4 | 1.5 |
| P38 | Scrum Master | Info Tech | 3 | 1 |
| P39 | Manager | Info Tech | 13 | 8 |
| P40 | Team Lead | Networking | 4 | 3 |
| P41 | Quality Assurance Lead | Networking | 2 | 2 |
| P42 | Scrum Master | Finance | 11 | 3 |
| P43 | Development Manager | Info Tech | 13 | 4 |
| P44 | Product Owner | Info Tech | 9 | 6 |
| P45 | Scrum Master | Info Tech | 20 | 12 |

**TEAM OBSERVATIONS**

| T# | P# | Team Size | Practices (N) | Duration (mins) |
|---|---|---|---|---|
| T1 | P22-P28 | 6-10 | SP (2), DSM (5), RT (1), TB (1), CR (1), BP (1) | 60, 10-15, 60, 120, 30, 30 |
| T2 | P38-P39 | 6-10 | SP (1), DSM (1) | 40, 10-15 |
| T3 | P40-P41 | 20-25 | DSM (5), QDR (1) | 20-25, 30 |
| T4 | P43-P44 | 20-25 | SA (1), DSM (1) | 115, 20-25 |
| T5 | NA | 10-15 | SP (1), RF (1), RT (1) | 90, 15, 45 |

Participant **P#**, **Domain** [Info Tech=Information Technology; IC Tech=Information and Communication Technologies], total experience **TX**, agile experience **AX**, Team **T#**, Observations count (**N**), Sprint planning **SP,** Daily stand-ups **DSM**, Retrospectives **RT,** Backlog prioritisation **BP,** Task breakdown **TB**, Sprint analysis **SA**, QA & design review **QDR**, Refinement **RF**

The Guide and Primer generally compensate (one provides information where the other is silent on the issue) and complement (one expands on what the other prescribes) each other but do not contradict each other.

### 3.1.2 In Practice

To understand what occurs in practice, we collected data through pre-interview questionnaires, semi-structured interviews, and observations. Participants were recruited by posting calls in popular agile meetup groups in New Zealand, Pakistan, and India and general posts on LinkedIn.

**Pre-interview questionnaires** were sent to each interviewee to gather basic demographic information about the participant, their team, and organization and information on their use of agile practices. The responses to these online questionnaires helped to focus the interviews. *Questions included: What is your total experience in the software industry (years)? What is your total agile experience (years)?*

To assess the frequency and maturity of their agile practices, we included the question: *Rate the frequency (never, rarely, occasionally, frequently, always) with which you perform the following agile practices*, with a list of the top 15 most common agile practices from the annual State of Agile survey [8]. All participants reported following Scrum practices such as sprint planning, daily stand-up meetings, and retrospectives, with varying frequencies.

**Semi-structured Interviews.** A total of 45 participants, seven from India (P1-P7), eight from Pakistan (P38-P44), one from the United Kingdom (P45), and twenty-nine from New Zealand (P8-P36) were interviewed. Interviews lasted between 30 and 60 minutes and were conducted either face-to-face (n=43) or via Skype (n=2). The demographics of the participants and teams' observations are summarized in Table 1. The participant numbers [P1-P45] and team numbers [T1-T5] are used to keep team and participant's anonymity as per the university's human ethics guidelines. The column header, role lists participant's primary roles in the team, e.g. product owner, scrum master, developer, tester etc. followed by the project domain, e.g. accounting, healthcare, finance; while the remaining columns list participant's years of total professional experience (TX), and agile experience (AX).

The authors collectively prepared the interview guide, conducted interviews [author1: n=38; author2: n=7) and analysed [all authors] to mitigate potential bias. All the interviews were recorded and transcribed for analysis either by the first author or third-party transcribers. Some of the questions asked during the interviews are: *1] What is the source of the business requirements for your project? 2] How does work item definition take place in your team? 3] How do you perform task breakdown in your team? 4] How and when does task allocation happen in your team? Can you share a few examples? 5) How and when does estimation take place in your team?*

**Observations** of five Scrum teams (two from New Zealand [T1, T5] and three [T2, T3, T4] in Pakistan) were conducted. We observed Scrum practices such as daily stand-up meetings, sprint planning, refinement sessions, breakdown sessions, retrospectives, and reviews. The second section of the table lists the teams' observations including the number, name, duration of the practices observed against the team number and the team size.

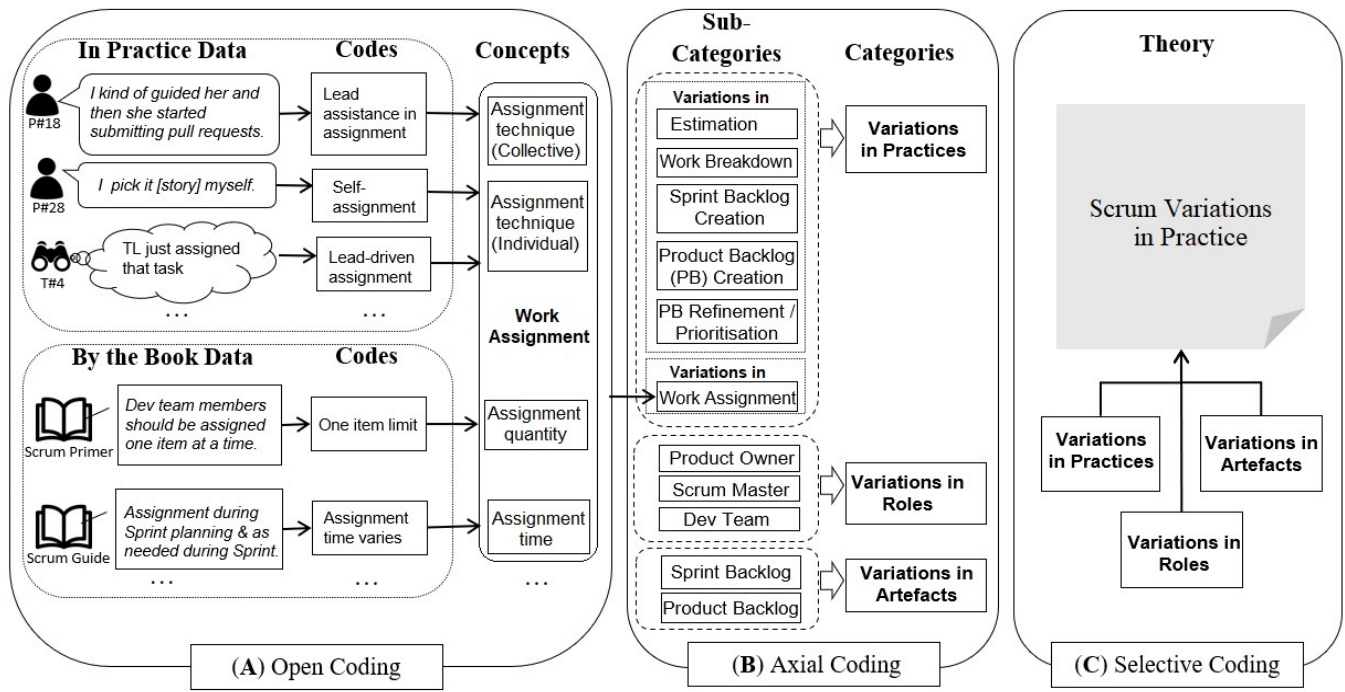

Fig. 1 Application of Strauss-Corbinian Grounded Theory steps: open coding (A), axial coding (B), and selective coding (C) leading to the theory of Scrum Variations in Practice.

These observations supplemented our understanding of the Scrum process, practices, strategies, and variations adopted by the team and corroborated our findings from the interview data. We also interviewed some members of the observed teams (all seven members of T1, 2 members of T2, 2 members of T3, and 2 members of T4). Interviewing them helped us to gather additional information or clarify any doubts recorded during the observations.

Data collection was performed in stages with interweaved rounds of data collection and analysis [25]. We continued collecting data as more and more variations in various Scrum practices kept being mentioned. The last rounds of interviews (P37-P45) and observation (T5) firmly indicated *theoretical saturation* as no new concepts, categories, or insights emerged.

### 3.2 Data Analysis

The *Strauss-Corbinian* version of GT includes three data analysis procedures: open, axial, and selective coding [25]. All these procedures and the emerging codes, concepts, sub-categories, and categories were mutually agreed upon through frequent and detailed discussions between the co-authors, including a GT expert, throughout the analysis. This resulted in further insights captured as memos [30]. All data such as transcripts, observation notes, artefacts, open codes, and memos were saved and processed using the NVivo data analysis software. The outcome of the study is a theory of Scrum variations in Practice with a set of categories of variations in practices, roles, and artefacts.

#### 3.2.1 Open Coding

We iteratively analyzed the interview transcripts and observation notes using open coding [25]. Fig 1 (A) illustrates the open coding and constant comparison procedures through an example, starting from the raw interview transcripts, observation notes and text from Scrum by the book. The two data sources, industrial data collected during the study (in practice) and the basic Scrum guides (by the book) were kept and analysed separately. Applying open coding on the raw data, **key points** were represented with short summary phrases and then further condensed into **codes** of 3-4 words each. As presented in the Fig 1 (A), *'lead assistance in assignment'* and *'self-assignment'* emerged as two different codes from the interview transcripts of two participants P18 and P28, and *'lead-driven assignment'* emerged as another code from the observation notes collected while observing the sprint planning of team T4. Through constant comparison, these and codes from other interview transcripts and observations were grouped to produce a higher level of abstraction, **concepts**, in this case, *'Work Assignment'*.

Open coding and constant comparison were also applied to the Scrum by the book as shown in Fig 1 (A). As an example, text from Scrum Primer resulted in code *'One item limit'* and similarly text from Scrum Guide resulted in code *'assignment time varies'*. These codes were grouped under two different concepts *'assignment quantity'* and *'assignment time'*. These and other assignment-related concepts (e.g. assignment techniques, assignment quantity) were grouped under a higher concept *'Work Assignment'*, shown in see Fig 1 (B). Similarly, we analyzed other practices of *estimations, breakdown, sprint backlog-creation, product backlog-creation,* and *refinement* and grouped them under respective higher concepts. This served as our basis for comparison to identify the variations between practice and by the book between these sub-categories. In this example, variations related to assignment found between the prac-tice data and by the book data led to the next level of abstraction, the sub-category, *'variations in work assignment'* (Section 4.2.3). Following the same procedure, we derived other sub-categories, such as *'variations in estimations'* (Section 4.2.1) and *'variations in breakdown'* (Section 4.2.2).

### 3.2.2 Axial Coding

Axial coding, a hallmark of the *Strauss-Corbinian GT*, is the *'process of relating subcategories to a category'* [25]. Using axial coding (see Fig 1.B), we identified relationships between our sub-categories and categories. This was driven by team discussions which involved activities such as referring to both by the book and in practice data iteratively for contextual details to identify correlations, drawing out the relationships on a whiteboard, and refining those relationships through discussions with further insights. These relationships evolved iteratively and retrospectively overtime. We related sub-categories to **categorie**s w.r.t. properties (techniques/granularity/when/units.) During this process, the 11 sub-categories were related to three main categories (see Fig 1.B). Practices-related variations (estimation, breakdown, assignment) were linked to category *'Variations in Scrum Practices'* (Section 4.2), roles-related variations mapped to the category *'Variations in Scrum Roles'* (Section 4.1). The associated variations in artefacts were linked to the category *'Variations in Scrum Artefacts'*, presented as part of Section 4.2.

### 3.2.3 Selective Coding

During selective coding (Fig 1.C), the sub-categories and categories derived from the open and axial coding were related to identify the **core-category** which explains the central phenomenon and builds the storyline or theory of our study [25] i.e. *a grounded theory of Scrum Variations in Practice.*

*Memos*, researcher-written notes exploring relationships in the emerging findings, helped to relate the categories and sub-categories to the core-category and to uncover the variations within and across teams, and between in practice and by the book data.

Finally, moving from description to conceptualisation [24], [25], we investigated the need for these variations in practice, from the collected data, presented as **rationales** in Section 4 of the paper. Guided by these, we defined our classification approach to refine our Grounded theory of *'Scrum Variations in Practice*': *standard, necessary, contextual*, and *clear deviation*, for practical and research use. These nuances (degrees) of variations emerged at the later stages of analysis while understanding the need for these variations. Not all variations occurred due to same rationales, some variations were based on the need, choice, context, while others arose from missing clarity in theory. Based on these observations, the second author proposed the nuanced *Scrum variations classification* approach, described in section 5.1. The first and second author discussed the classification approach using multiple examples from the underlying data and analysis. The first author then classified each variation evidenced in the underlying data using the new classification approach, while the third author reviewed the process and approach.

## 4 RESULTS

**RQ:** *How, when and why does Scrum practice vary from Scrum by the book?*

We found variations between Scrum by the book and in practice across three categories:
- *Variatons in Scrum Roles* (section 4.1)
- *Variations in Scrum Practices* (section 4.2)
- *Variations in Scrum Artefacts* (discussed with roles and practices)

The *variations in Scrum project management practices* span across: estimation, breakdown, assignment, sprint backlog creation, product backlog creation, and product backlog refinement/prioritization.

Section 4 details the *how, when,* and *why* variations occur, and Table 2 presents a summarized overview of the variations, including rationales (*why*).

In this section, we describe the main components of our findings: *variations in Scrum roles* (Section 4.1), *variations in Scrum practices* (Section 4.2), and associated *variations in artefacts*, embedded as descriptions within the other two sections.

In presenting the results, we first describe what Scrum prescribes **by the book** followed by what we found **in practice**, describing the practices that were in line with Scrum by the book and the many variations we discovered along with their rationales.

Throughout this section, we include several original quotes from the interviews and draw on observations to support our descriptions and the verifiability of our work. While this is not a quantitative study, we use some terms throughout the text to indicate the extent of the prevalence of practices: 'few' refer to less than 25%, 'many or majority' refer to greater than 70%, and other cases are referred to by terms such as some, frequently, often or very often.

### 4.1 Variations in Scrum Roles

#### 4.1.1 Product Owner

**By the Book:** As per the Scrum Primer [2], the PO represents the customer and is responsible for translating the desired product features into a prioritised list of items. The PO acts as a bridge between the development team and stakeholders, such as customers. For internal projects, the PO and customer are often the same person [2].

**In Practice:** Most teams had a dedicated PO, which aligns with Scrum by the book. Some organizations had more than one Scrum team that shared the same PO.

*"We have 2 POs for our entire product with different portfolios."* P#37

While theory dictates having a single person acting as PO, some POs had an extended support team of business analysts to share some of the PO duties. Conversely, other teams had the PO perform additional duties of other roles.

*"The PO is also the technical manager and business analyst"* P#11

A few teams working in service-based domains had the customer working as the PO.

*"In most of our cases our PO is actually the client because we are a services company, we are not building our own products." P#39*

### 4.1.2 Scrum Master

**By the Book:** The Scrum Master (SM) helps the team stay organized and ensures Scrum method adherence. The SM coaches the team and facilitates issue resolutions. The SM is usually a dedicated role, but smaller teams may have a team member doubling up on this role [2]. Prerequisite background skills for the SM are not specified.

**In Practice:** Most of the teams had a dedicated SM. In some cases, teams had a single, shared SM. A few teams had no SM at all because they considered themselves mature enough to not need one or the previous SM had left. It was also common for an experienced member, e.g. team lead, to serve as the SM on a rotational basis.

*"We had a volunteer Scrum Master within the team to act for the two weeks' time." P#36*

### 4.1.3 Development Team

**By the Book:** The development team is a group of seven (plus or minus two) self-managed, autonomous team members who possess the expertise necessary to deliver a potentially shippable product. Scrum team members are encouraged to be cross-functional. There are no designated roles or titles such as tester, business analyst, or programmer [1], [2].

**In Practice:** In contrast to what Scrum by the book states, there were very few cross-functional teams. Most teams had specialists in specific domains or areas, such as front-end or back-end development, due to their prior experience. Additionally, members often had specific designated roles such as testers, developers, and business analysts, which does not comply with Scrum by the book.

*"Our team make-up is we have four developers, two testers and a BA and a PO in our team." P#12*

While Scrum by the book encourages cross-functional teams, our results confirm it is not uncommon that a team of specialists practice Scrum. It is seen that specialists do not turn into cross-functional teams instantly, it happens over time due to factors such as less visible and immediate benefits, and support of management, team, and individuals.

### 4.2 Variations in Scrum Practices

We now present the identified variations in Scrum project management practices: estimation, breakdown, assignment, sprint backlog creation, product backlog creation, and product backlog refinement/prioritization.

Table 2 presents a summarized overview of the variations in the practices along with the associated roles and artefacts. The first column captures the Scrum project management practices. Each of these spans three rows. The first row (without shading) represents By the Book (**B**), the second light grey row lists what actually occurs in practice (**P**), and third the dark grey row lists the rationales behind the variations (**R**). We number each rationale as [R#], going from R1-R20, and use this notation throughout the results to map back to the summary in Table 2.

#### 4.2.1 Estimation

**By the Book:** Estimation involves predicting the effort required to carry out a work item. Scrum teams are meant to estimate collectively. Estimates can be measured in different ways, e.g. in person hours/days or story points [31]. The Scrum Guide specifies that Product Backlog items must have an estimate, but it does not impose any particular estimation technique or prescribe when estimation must be done.

**In Practice:** There are variations in *who* does estimation, *how* it is done, and *when* it is done.

**Individual Estimation:** Contrary to Scrum by the book, members often estimated work items individually. **Team leads or an experienced developer** used their domain knowledge and experience to ensure **accurate estimates** [R4].

*"We [team] rely on our technical leads for estimates." P#37*

Sometimes, the individual estimation was prompted by the PO who asked an individual team member for an estimate due to their expertise (rationale [R3]).

*"Most of the time the PO goes, 'Hey, can you have a look at this, and come back to me with how long that's going to take'." P#35*

Individual estimation was also seen in a case where an **influential PO** made the estimates themselves (rationale [R3] in Table 2).

**Collective Estimation:** Collective estimation was observed in most of the Scrum teams. As noted during an observation of a sprint planning session, the estimates were proposed, discussed, and evaluated collectively by the **entire team**. This also happened **in pairs** (typically the lead and a developer). In many cases, the team lead made the final decision after the collective discussions.

*"Dev [Development] lead sets the estimation in hours after discussing with the developer." P#44*

When the entire team was involved, the SM (sometimes) and PO (almost always) also participated. The involvement of the PO helped the team understand work items and the PO to set reasonable expectations and priorities for future sprints (rationale [R2] and [R3] in Table 2). But, involving all members was also reported as ineffective resource utilization (rationale [R3]) (P#43).

**Estimation Techniques and Units:** Scrum teams followed many different techniques, such as Planning Poker using fingers or cards. The units of measurement also varied, including story points, hours, and t-shirt sizes. The techniques and units were selected based on team and individual preferences (rationale [R1]) as indicated in the listed quote where SM expresses dislike towards planning poker and prefers t-shirt sizes instead.

*"I [Scrum Master] hate that [Poker], that's meaningless, it's just a number... how long it takes, so that's what the PO wants to know. So, I prefer my team to give a rough estimation in small, medium and large." P#19*

**Estimation Levels:** Many Scrum teams did estimations at two levels: for the Product and Sprint backlog items. Interestingly, teams used different units (e.g. points for PB and hours for SB) as per their preferences (rationale [R1]).

TABLE 2 SUMMARY OF VARIATIONS IN SCRUM PRACTICE

| PM Area | B/P/R | Practice | Roles | Artefacts |
|---|---|---|---|---|
| ESTIMATION | B | *Techniques:* Not prescribed, relative size<br>*Units:* Not prescribed, story points<br>*When:* Not prescribed, before Sprint | Development team (collective) | Product backlog |
| | P | *Techniques:* Varied (e.g. planning poker) [R1]<br>*Units:* Varied (e.g. story points, hours) [R1]<br>*When (PB):* Refinement session [R2]<br>*When (SB):* before/during Sprint planning<br>*When (changes):* any time | *Individual* estimation [R3]<br>Team Lead, developer (domain expert) [R4], or PO<br>*Collective* estimation<br>▪ Development team + SM and PO [R5]<br>▪ Pair (Team Lead + assigned dev) | Product backlog<br>Sprint backlog [R6] |
| | R | [R1] team and individual preferences<br>[R2] more accurate SB estimates, help PO set priorities, greater autonomy | [R3] more effective resource utilization<br>[R4] accurate estimates<br>[R5] increases PO clarity of priorities and team understanding of user perspective<br>[R6] individual accountability, manageable workload | |
| BREAKDOWN | B | *Techniques:* not prescribed<br>*Granularity:* stories -> tasks<br>*When:* during Sprint planning (current Sprint)<br>*Units:* one day or less | Development team (collective) | Sprint backlog |
| | P | *Techniques* vertical [R9], horizontal<br>*Granularity:* stories -> tasks or sub-tasks [R7], no breakdown<br>*When:* Sprint planning, during Sprint, never<br>*Units:* one day or less, max points per task | *Individual* work breakdown [R8]<br>*Collective*<br>▪ Development team + SM {additional people okay}<br>▪ Pair [R14] (same roles; same or different product area) | Sprint backlog |
| | R | [R7] better understanding, involvement | [R8] expertise, domain knowledge leads to better breakdown<br>[R9] earlier customer delivery and feedback | |
| ASSIGNMENT | B | *Techniques:* Self-organize, volunteer<br>*When:* Sprint planning, during Sprint<br>*Quantity:* one assigned item per team member | Development team (individually volunteer) | Sprint backlog |
| | P | *Techniques:* Self-organize [R9], Manager/TL assigned<br>*When:* Sprint planning, daily standups, during Sprint<br>*Quantity:* assigned item(s) per team member [R1]<br>*Techniques [selection criteria]:* ad-hoc [R10], dedicated [R11] | *Individual* assignment<br>▪ Dev team member self-assigns<br>▪ Lead/manager assigns [R13]<br>*Collective* assignment [R14]<br>▪ Dev team collectively<br>▪ Pairs (Team lead or SM with dev) | Sprint backlog |
| | R | [R10] many factors, e.g. interests and opportunity to learn<br>[R11] faster completion, individual accountability<br>[R12] empowerment, autonomy, learning, manager time saved | [R13] team lack experience or domain knowledge, urgent work<br>[R14] shared accountability, knowledge sharing, helps new or inexperienced team members | |
| SPRINT BACKLOG CREATION | B | *When:* Sprint planning<br>*Goal setting:* one Sprint goal (optional)<br>*Order:* define goal then select items<br>*Quantity:* based on velocity | Development team (collective) | Sprint backlog |
| | P | *When:* Sprint planning<br>*Goal setting:* one goal; multiple goals [R15]; no goal<br>*Order:* define goal then select items to fit, pull first then define goal(s) to fit<br>*Quantity:* velocity, velocity + stretch tasks | *Individual* sprint backlog creation (PO [16], a business consultant, a project manager, or the client)<br>*Collective*<br>▪ dev team + SM<br>▪ PO or customer + dev team rep | Sprint backlog |
| | R | [R15] hard to map items to one goal | [R16] higher visibility, knows what they want from Sprint | |
| PRODUCT BACKLOG CREATION | B | *Content:* ordered list of features, functions, requirements, enhancements, and fixes with description, order, estimate, and value<br>*Type of work items:* technical; user stories or other<br>*Tools:* not prescribed | PO (individual) | Product backlog |
| | P | *Content:* ordered; un-ordered; semi-ordered<br>*Type of work items:* technical and non-technical; epics, features, stories, tasks, incidents, tickets, bugs, and spikes [R17]<br>*Tools:* online project management tool like Jira, and Team Foundation Server | Varies. PO, clients, end-users, support team [R18] | Product backlog |
| | R | [R17] team preferences | [R18] organisational structure | |
| PRODUCT BACKLOG REFINEMENT / PRIORITISATION | B | *When (Prioritization):* refinement session<br>*What (Prioritization):* All PB items<br>*When (Refinement):* refinement session | *Prioritization*: PO (individual)<br>*Refinement*: Development team + SM + PO (collective) | Product backlog |
| | P | *When (Prioritization):* when adding to PB; during refinement; during Sprint planning<br>*What (Prioritisation):* all /some/no PB items [R20]<br>*When (Refinement):* Sprint planning (current Sprint), refinement session (future Sprint) | *Prioritisation*<br>▪ PO or a Business Analyst or Business consultant<br>▪ Development team (collective)<br>*Refinement*<br>▪ Development team (collective)<br>PO + dev team representative [R19] | Product backlog |
| | R | [R19] scope change | [R20] contextual factors | |

Non-shaded (white) rows summarise what Scrum **by the Book (B)** states about the Scrum practices, roles and artefacts, light grey rows list the variations **in practice (P)**, and dark grey rows list the **rationales** behind the variations **(R)**. Each rationale is numbered [R#] and is used to map to the relevant variation in practice (in the P row) in superscript [R#].

*Product backlog estimation* was conducted during refinement sessions, mostly using story points. Teams that estimated product backlogs well in advance were seen to make more accurate estimates for the sprint backlog, making it easier for the PO to set priorities (rationale [R2]). However, not all teams made early estimations.

*Sprint backlog estimation* was typically conducted before or during Sprint planning. Most teams collectively estimated tasks using hours as the unit. Teams that estimated Sprint backlog items were more likely to create a manageable Sprint workload and displayed individual accountability (rationale [R6]). However, not all teams estimated the Sprint backlog items; some simply pulled the estimated product backlog items directly onto the Sprint backlog.

**Changes to estimations:** Estimates for Product or Sprint backlog items can be increased or decreased at any time with reason. In one case, during a Sprint backlog refinement session, we observed a team re-estimating several product backlog items because not all team members had been involved in the original estimation (rationale [R7]).

Another team re-estimated a couple of the Product Backlog items when they discovered a scope change (rationale [R19]). When estimations changed, the team discussed this in the daily stand-ups and work assignments were modified accordingly.

*"When you estimate task it's not a line cut in stone. Obviously, you have some room in that…no one will stop you to update the related estimations if needed [In Jira]." P#9*

### 4.2.2 Work Breakdown

**By the Book:** The Product backlog has items of varying sizes and complexities. During the Sprint planning meeting, the team decomposes the highest priority user stories into individual tasks 'to units of one day or less' [1]. The entire development team should participate.

**In Practice:** There are variations in who does the breakdown, how and/or when it is done, and the granularity.

**Individual Breakdown:** Contrary to Scrum by the book, team members performed the breakdown independently in some cases, relying on their own expertise (rationale [R8]). This usually happened during the second part of the sprint planning after the team had selected the stories for the sprint.

*"Then, we pick the story [second part of sprint planning], and then we'll break it down into the tasks ourselves [individually] what we think we need to do [for the entire sprint]." P#24*

**Collective Breakdown** was done by most teams by the entire team or in pairs. This was reported to improve shared understanding and collective ownership of tasks, especially for inexperienced members (rationale [R14]). When the entire team participated, breakdown was performed through discussions in the Sprint planning meeting. Either the SM recorded all tasks or members recorded their own tasks using post-its on a physical board or electronically.

*"Everyone writes a task ... one person typing at a time, and we just pass the keyboard around..." P#11*

*"We [developers] would do that during that session, while we were discussing the solution, the Scrum master would be sitting and typing tasks." P#18*

When a pair performed task breakdown, they would have the same role (e.g. two testers). However, the pair did not always work on the same product area, ensuring different perspectives were considered and knowledge transfer opportunities (rationale [R14]).

**Breakdown Techniques:** Teams broke down items either horizontally or vertically.

*Horizontal Breakdown* involves breaking down stories by the type of work required or the components that are involved (e.g. all the User Interface or database work).

*"Let's finish all the infrastructure for the project and then let's do all the backend, and then let's do all the front-end..." P#21*

*Vertical Breakdown* breaks down work items across functional layers so that new functionality can be delivered to the customer as early as possible (rationale [R9]).

*"So, to deliver value we should do as part of the story a bit of frontend, a bit of the backend, so that we can go and deliver something to the customer ASAP." P#21*

**Level of Granularity:** For most teams, new features were created as user stories (high granularity) and everything else, such as enhancements and bugs, were created as tasks (low granularity). During breakdown, teams decomposed user stories into tasks.

*"Usually the user story is what we [team] take in and we break down into multiple tasks." P#21*

Detailed breakdown helped the team members better understand the tasks during implementation (rationale [R7]). In line with Scrum by the book, teams reported breaking down the bigger stories first, aiming for some maximum number of points per task.

*"…if a story is bigger than eight points, then it's probably too big and we should try to break it down if we can." P#20*

However, some teams did not perform work breakdown at all treating the work item as a story throughout the development process (rationale [R1]).

*"We don't create separate tasks, it's just a story which covers development and testing work." P#32*

The level of granularity is also influenced by the experience of the team members. New teams were seen to use an overly detailed breakdown. For example:

*"reproduce the bug, fix the bug, and verify that it's fixed' this [level of breakdown] is an indicator of less experience". P#23*

On the other hand, mature teams may not need as detailed of a breakdown:

*"It was obvious that building an API would cover writing an endpoint, refactoring existing code, integrating the database change". P#21*

### 4.2.3 Work Assignment

**By the Book:** The Scrum team is meant to self-organise to carry out work assignments during the Sprint planning meeting and throughout the Sprint. Scrum by the book encourages people to volunteer for tasks, one at a time, based on business value. It also encourages selecting tasks that promote learning (e.g. by pairing with a specialist to work on something they are not skilled at).

**In Practice:** We observed a wide range of variations around who, when, how much, and how assignment occurs.

**Collective Assignment:** was practiced to support new or less-experienced members (rationale [R14]). Members were seen collaborating, offering help, and negotiating with each other while making assignment decisions. We observed a developer ask another developer during a daily stand-up to re-assign a task due to some unexpected technical issues. Such transparency held team members accountable collectively and supported knowledge sharing.

*"During the sprint, we [team] see if some tickets need to be shifted around and shuffled, that's on us." P#40*

*"If he [new member] can't decide which ones to choose… then team members will advise him saying, 'oh, try this one!'." P#30*

**Individual Assignment:** Many teams practiced self-assignment because this was recommended by Scrum by the book and was seen to encourage empowerment, autonomy, and learning, and minimized the time managers spent on assigning (rationale [R12]). However, individual work assignment in Scrum teams also happened through the manager or the team lead. This often happened when team members were less experienced or lacked domain knowledge (rationale [R13]). Other factors, such as urgency of the task, also caused managers or leads to directly assign work to the team members (rationale [R13]). Even teams practicing self-assignment had instances when work was assigned to them by the manager or lead.

*"there are urgent stuff that gets put onto my desk." P#22*

**Assignment Time:** Teams assigned tasks at different occasions during the Sprint including during the Sprint planning, during the daily stand-ups, or in an ad-hoc manner at any time during the Sprint. When assignment happened during the Sprint Planning meeting, it was in the presence of the development team, Scrum Master and PO or Technical Manager. If it happened during the daily stand-up, then the PO was not present which indicates that the presence of PO did not affect the self-assignment decision. When it happened ad-hoc during the Sprint, it usually only involved one or two team members, who recorded the assignment on the physical board or in the digital project management tool (e.g. JIRA).

*"Basically, what will happen is, when someone decides to pick up a task, they'll go to our physical board, and they'll move it to make it in progress. And they'll start work." P#20*

**Assignment Quantity:** Many teams followed Scrum by the book and volunteered for one task at a time [2]:

*"…when someone comes free, they'll just look down from the top, and go, okay, this is the next task that needs to be done" P#11*

But as a variation, it was not uncommon for teams to assign multiple work items to each member. Their complexity, relevance, and dependency influenced the number of items being assigned (rationale [R17]).

*"It depends on the complexity of the feature or the item or whatever it is, the task. They'll [developers] pick more than one, we'll [testers] pick more than one." P#32*

**Assignment Techniques:** Items were assigned in either an ad hoc or dedicated way. When ad hoc, members picked up any task from the Sprint backlog based on their interests, roles, expertise, opportunity to learn, or other factors (rationale [R10]). The dedicated technique meant team members were dedicated to finishing a user story and picked up tasks related to only that user story (rationale [R11]). Multiple team members could be dedicated to the same user story (e.g., we observed a developer working on the functionality while a tester worked on the associated test cases). This technique was more common in less cross-functional teams.

### 4.2.4 Sprint Backlog (SB) Creation

**By the Book:** During the Sprint Planning meeting, the Scrum team defines a sprint goal to set the objective of the Sprint and commits to a list of selected product backlog items, which becomes the Sprint backlog. The team should collectively pull and commit to these Sprint backlog items.

**In Practice:** We observed variations in creating the Sprint backlog and the Sprint goals.

**Individual creation:** In contrast to Scrum by the book, the entire team was not always involved in creating the Sprint backlog. We observed it being done by a single person: the PO, business consultant, project manager, or the client (rationale [R16]) due to their higher visibility of Sprint goals.

*"Our lead PO selects these tasks … for the sprint." P#44*

**Collective creation:** We observed cases where the team collectively selected the sprint items as suggested by the book. Even in these cases, due to a higher level of visibility and domain experience, the Team lead finalised the Sprint backlog (rationale [R16]).

*"I [team lead] decide them [Sprint items], coz I have greater visibility so I basically sit down with the team and then we prioritise the tickets like this feature needs to be implemented before that, or something like that." P#40*

Sometimes a team member would assist the PO or client in selecting the items for the Sprint backlog.

*"We [business and team representative] do a compilation of all the requirements received from customers, and internally from within the organisation. After prioritisation and triaging internally, we come up with a Sprint backlog." P#30*

**Quantity:** The number of items selected for the Sprint backlog was often based on the team's velocity, selecting the number of tasks expected to be completed in the Sprint. However, we also saw cases where the Sprint backlog had *'stretch tasks'* that were not expected but could be completed if others were finished (rationale [R16]).

**Sprint Goal:** In line with Scrum by the book, most teams set *a single, specific Sprint goal* during Sprint planning, facilitating their selection of the Sprint items. For example, a Sprint goal was to release a specific feature. With this goal in mind, the team pulled the stories aligned with the goal from the Product backlog.

*"The new thing we recently introduced [after X years] is we [PO and team] try to come up with a sprint goal and then pull the features which aligned with our sprint goals." P#37*

Other teams had *multiple, often unrelated, goals* or *sub-goals* where teams were often working on many features and created goals to fit the items on the Sprint backlog (rationale [R15]), rather than selecting the backlog items to fit the goal.

Some teams (T2, T3) had *no Sprint goal* due to lack of understanding on its purpose or difficulty in finding a common purpose due to scattered priorities (rationale [R15]).

### 4.2.5 Product Backlog (PB) Creation

**By the Book:** Keeping in mind the needs of the stakeholders and the business strategy, the PO is responsible to define the product features as PB items. The PB is a prioritized list of work items with varying sizes and details, but they are usually vague high-level descriptions of features in the form of user stories or use cases.

**In Practice:** We observed variations in who creates the PB items, type of work items, and tools used.

**Who:** In addition to the PO, backlog items came from a *variety of sources*: clients, end-users, and support team.

*"Generally, we get requirements directly from our client(s) who use our product, we also get ideas during the demonstration of our product to other prospects…strategic requirements too…proposed by COO… [and others] suggested by team or reported by the client or the QA team." P#44*

In some teams, PB items were created by someone other than the PO due to organization structure (rationale [R18]). For example, the Support Team could create a 'support ticket'. The team would then investigate if it described a bug, feature request, or enhancement and add it to the PB.

**Type of work items:** In addition to new features, the PB included both *technical work* (e.g. enhancing/maintaining features, fixing bugs, migrating data, configuring environments, reducing technical debt, and providing technical support to other teams) and *non-technical work* (such as creating user guides, conducting feasibility studies, preparing demos, and coordinating with other teams).

These work items were included on the PB as epics, features, stories, tasks, incidents, tickets, bugs, and spikes. The type of items varied across teams (rationale [R17]). Some teams tracked all work items as tickets or incidents irrespective of whether it was a bug, enhancement, or a new feature. Other teams reported treating all items as features represented as user stories.

*"If it's a bug then that is treated the same as a new feature [in the form of stories]." P#11*

**Tools:** Most teams use an online project management tool to host their PB items. Jira, Team Foundation Server, YouTrack, and GitLab were commonly used (rationales [R17] and [R18]).

### 4.2.6 Product Backlog Refinement and Prioritisation

**By the Book:** The PO is responsible for prioritising the items on the PB. The highest priority items should be refined *(or groomed)* with additional details to allow the team to execute them. Refinement is done collaboratively with the team, SM, and PO.

**In Practice:** We observed variations in who did the refinement and prioritisation and when (if) it occurred.

**Prioritisation:** The PO or a Business Analyst was often responsible for prioritisation, but the team also prioritised in some cases. We observed that the PB was not always prioritized or items were *missing* priorities. During a Sprint planning meeting, we observed one team, with a non-prioritized PB, prioritise items as they added them to the Sprint backlog.

Prioritisation was based on a variety of contextual factors, such as users' requests, severity of a particular issue or new requirements, or the estimated impact of the work item (rationale [R17]).

**Refinement** was typically done *collectively* through discussions with the entire team, in line with Scrum by the book. This enabled the team to provide the technical perspective while the PO provided the business perspective. However, sometimes it involved only the PO and the team lead to represent the team to address any scope change, or because some refinements did not need the entire team (rationales [R17] and [R19]). Refinement typically occurred in the first part of the Sprint planning meeting or in a refinement session.

## 5 DISCUSSION

### 5.1 A Nuanced Classification Approach

Based on the evidence around variations seen in practice (study data) and in careful comparison to the prescribed *Scrum by the book* (the Scrum Guide and Primer), we classify variations as:

1. *standard variations,* variations allowed by the book,
2. *necessary variations,* variations created in practice to address vagueness or ambiguity in Scrum by the book,
3. *contextual variations,* temporary and/or infrequent *justified* variations contradicting Scrum by the book,
4. *clear deviations,* ongoing or frequent *unjustified* variations contradicting Scrum by the book.

We present these classifications using examples from the study that had enough supporting details and evidence. We do not classify all variations observed (e.g. in Table 2), because not all have enough contextual information to warrant confident classification.

**Standard Variations** are specific variations already mentioned in Scrum by the book as optional implementation pathways. An example of this is *assignment time*. In Scrum by the book, the Guide states that work assignment can occur *"both during sprint planning and as needed throughout the sprint"*, a notion supported by the Primer. In practice, assignment occurred during sprint planning, daily standup, and on an ad-hoc basis through the sprint, following the variations allowed by the book. In other words, standard variations were observed in practice.

Another example of standard variations is *estimation techniques*. The word *'estimate'*, in the context of estimation, appears 9 times in the Guide and 37 times in the Primer. The Guide does not prescribe *how* to estimate but the Primer compensates by recommending *"relative size"* as a guideline and *"story points"* and *"hours"* as concrete examples of allowed variations. In practice, estimation was practiced as per standard variations, i.e. using story points and hours as allowed by the book, and also using t-shirt sizes (small, medium, large, extra-large), which although not mentioned as a specific example in the Primer follows the guideline around using a *"relative size"* measure.

**Necessary Variations** are variations that are created to address vagueness or ambiguity in Scrum by the book. An example of a necessary variation is Scrum teams *adapting the order* of the project management practices. While Scrum by the book (both Guide and Primer) refers to breakdown, estimation, and assignment, it is unclear what order they are meant to occur in, or whether a particular order is preferred. Such ambiguity necessitates variations in practice. Some teams estimated items before breakdown and assignment (e.g. T1, T5). Others performed assignment before estimation and breakdown (e.g. T2).

Another example of necessary variations is *refinement*. The Guide leaves the implementation of the refinement practice to the team, stating *"Scrum teams decides how and when refinement is done"*. In practice, teams held specific refinement sessions before sprint planning in some cases (T5) and during analysis sessions in others (T4).

**Contextual Variations** contradict what Scrum by the book prescribes, *justified* by rationales covering practical constraints and contextual factors, resorted to on a temporary or infrequent basis, typically with the intention to align with Scrum by the book over time.

An example of a contextual variation is *assignment quantity and technique*. The Guide does not prescribe how many items should be assigned to individuals but is compensated by the Primer, which clearly states *"volunteer one task at a time…that will on purpose involve learning"*. The latter part of the statement supports the cross-functional teams' concept, also promoted by the Guide. In practice, teams (T1, T5) selected multiple items during sprint planning based on individual expertise and specialisation (as opposed to cross-functionality) contradicting Scrum by the book. However, this was justified in case(s) where: the team was still transitioning into Scrum and their cross-functionality had not matured. We know this was temporary because these teams were also observed practicing learning-led self-assignment on a smaller scale.

Another example of contextual variation is *work assignment*. Both the Guide and the Primer recommend *"self-assignment"* and explicitly discourage delegation. In practice, the team lead practices delegation or direct assignment during early stages of onboarding novice members (P#18) or the manager delegates urgent high priority items to the most skilled person for faster delivery every once in a while (T1 or P#22). Both cases represent temporary contradictions to Scrum by the book with justifications.

**Clear Deviations** are variations that contradict what Scrum by the book clearly prescribes, *not justified* by rationales, and practiced on a frequent or on-going basis, typically with no intention to align with Scrum by the book over time. An example of a clear deviation is **team lead-driven assignment on a regular and/or permanent basis** with no effort to transition closer to self-assignment (T3). Another example of clear deviation is the *PO/Business Consultant/Project Manager **deciding how much and what work the team will deliver during the Sprint*** in practice (T4, T3). This is contrary to Scrum Primer stating *'Team decides how much work it will complete, rather than having it assigned to them by the Product Owner'*.

Clear deviations likely stem from misunderstanding of Scrum by the book or as remnants of traditional software development mindsets and can be considered misuse or abuse depending on intention.

Variations to *Scrum by the book* are inevitable. Method tailoring, adaptations, and deviations of software development methods have been acknowledged for the past two decades [11, 12, 13, 16, 40, 43]. However, this prior work did not consider the classification of variations, considering any variation as misuse or abuse. From our findings, we show that there are different types of variations including some that are required and necessary. We present a nuanced Scrum variations classification approach. Future work can extend our nuanced approach to differentiate when these Scrum variations can facilitate different settings e.g. extending Scrum to scaled or distributed software teams or merging with other agile methods leading to hybrids [43].

> **A Nuanced Scrum Variations Classification Approach**
>
> Variations to *Scrum by the book* are inevitable. Not all variations are process misuse or abuse. Our nuanced Scrum variations classification approach explains variations in practice as:
> - *standard variations*, variations allowed by the book
> - *necessary variations*, variations created in practice to address vagueness or ambiguity in Scrum by the book
> - *contextual variations*, temporary and/or infrequent *justified* variations contradicting Scrum by the book, and
> - *clear deviations,* ongoing or frequent *unjustified* variations contradicting Scrum by the book, excuses for poor implementation.
>
> Our classification approach can be extended to make sense of variations in other Scrum practices and potentially in other agile methods and practice frameworks.

### 5.2 Recommendations for Scrum Practitioners

1. Use of standard variations are in line with Scrum by the book and within the range of allowed variations.
2. Because of vagueness or ambiguity in Scrum by the book, practitioners must apply necessary variations. *Necessary variations do not constitute misuse or abuse.*
3. Contextual variations are applied temporarily to address contextual constraints, e.g. while a team transitions into Scrum, with conscious effort to move closer in line with Scrum by the book over time. As such, *contextual variations are not misuse or abuse of Scrum.*
4. *Clear deviations are juxtaposed to the essence and fundamentals of Scrum* and are excuses for not implementing Scrum by the book, same as 'ScrumButt'.

Based on our own comparative analysis between Scrum by the book and in practice and previous related work [19], we propose that part of Scrum's sustained growth in industrial practice over the years can be attributed to two factors: the light-weight and flexible nature of its seminal guides [1], [2] such that Scrum by the book is neither entirely vague nor completely prescriptive; and the Scrum variations in practice, enabling real-world software teams to tailor it to their needs. The flipside of these same factors may explain in part XP's steady decline in industrial popularity [1], [8], [35], i.e. the relatively elaborate XP guidelines [36] and its documented resistance to tailoring [19]. Based on these observations, we recommend that both practitioners and researchers **avoid hastening to label all variations as deviations**, instead use our nuanced classification approach to make sense of Scrum variations.

### 5.3 Limitations and Verifiability

A Grounded Theory study does not claim generalization, rather produces a mid-ranged theory applicable to the contexts studied [30], [32]. Our data collection does not represent the entire international agile community and is limited to agile practitioners who responded to our call for participation. The details of these participants, their companies, and third-party clients have been kept confidential as per the human ethics guidelines governing this study.

Throughout the study, the data collection and analysis procedures, emerging codes, and insights were collaboratively discussed, debated, and finalized by all authors to overcome any potential biases. We propose our variations classifications can be extended to apply more widely to other aspects of Scrum, beyond project management practices, and potentially to other agile methods, however, this remains to be validated in practice. We hope future studies can use, validate, and extend our classifications. The study focus is key project management practices such as planning (includes. estimation, breakdown, SB/PB creation, prioritization) and assignment so variations in practicing retrospectives and sprint reviews or the quality management, design or implementation aspects are out of scope. However, retrospectives could often bring up assignment and planning issues, so we included them as part of our data collection while interviewing participants and observing teams', but our findings did not identify significant variations focusing project management practices. We believe Scrum variations towards quality management, design and implementation is another facet of Scrum and could be included in future studies.

The verifiability of a grounded theory (outcome) can be derived from the robustness of the GT (method) as evidenced from the description of its application [28], [29], [32]. To achieve this, we have described our application of the Strauss-Corbinian GT method in substantial detail (Section 3 and Fig 1) and included original quotes from the underlying data in our description of the findings (Section 4). In doing so, we have demonstrated how our theory fulfills the GT evaluation criteria: (a) the categories derived *fit* the underlying data (see Fig 1 and Table 2), (b) the theory *works* in that it explains the main concerns of the participants (practicing Scrum within real-world constraints) while answering the research question, (c) it has *relevance* for the agile practice and research communities, and (d) is *modifiable* through future studies [32].

## 6　Conclusion

Scrum is a popular agile method that can be difficult to implement by the book since it does not prescribe the *'how'* for many of its practices, roles, and artefacts. Labeling all variations as misuse, abuse, and deviations displays oversight of the vagueness inherent in the fundamental Scrum guidelines and of real-world challenges and constraints practitioners face. Between the two extremes, Scrum by the book and ScrumButts, a variety of variations exist and may be necessary in real-world software projects.

Our theory describes variations in Scrum practices, roles, and artefacts and their underlying rationales. Through empirical evidence of Scrum variations based on extensive GT analysis of Scrum by the book (i.e. Guide and Primer) and Practice (i.e. 45 interviews and 5 observations), we introduce a nuanced approach to understanding variations. Variations are classified as *standard* (listed in Scrum by the book), *necessary* (required due to vagueness or ambiguity in Scrum by the book), *contextual* (temporary or infrequent justified variations contradicting Scrum by the book), and *clear deviations* (ongoing or frequent unjustified variations contradicting Scrum by the book). Clear deviations are misuse or abuse, same as ScrumButt.

We believe acknowledging and understanding the need and use of these variations will help Scrum by the book work in practice. Our findings and classification approach lay the foundations for future research. Future studies can investigate the impact of these variations on productivity and quality and extend our variation classifications.

### Acknowledgment

We would like to thank all the study participants. This study was conducted with approval from the Human Participants Ethics Committee at the University of Auckland.

## AUTHOR BIOGRAPHIES

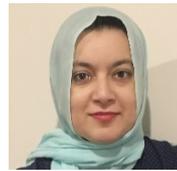

Zainab Masood is currently pursuing her doctoral degree at the University of Auckland (Electrical, Computer, and Software Engineering), New Zealand. Her research interests include agile software development, software testing and quality assurance, and human aspects of software engineering.

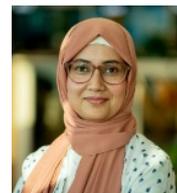

Rashina Hoda is an Associate Dean (Academic Workforce) and an Associate Professor in software engineering at the Faculty of Information Technology at Monash University. Her research focuses on human-centred software engineering, agile software development, and grounded theory. She serves on the IEEE TSE reviewer board and IEEE Software advisory panel, and Journal of Systems and Software editorial board. Rashina is currently writing a book on Grounded Theory for Software Engineering. More: www.rashina.com

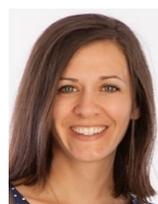

Kelly Blincoe is a Senior Lecturer at the University of Auckland's Department of Electrical, Computer, and Software Engineering. Her research is mainly in the human aspects of software engineering with a focus on collaborative software development and software requirements. She currently serves on the editorial board of the Empirical Software Engineering Journal and the Journal of Systems and Software. She is also on the Executive Board of Software Innovation New Zealand. For more information visit: kblincoe.github.io